\documentclass{aastex}
\usepackage{emulateapj5,psfig}

\newcommand{\etal}{{\it et al. }}

\newcommand{\solar}{\ifmmode_{\mathord\odot}}

\newcommand{\beq}{\begin{equation}}
\newcommand{\eeq}{\end{equation}}

\lefthead{HINKLEY \etal}
\righthead{Optical$-$NIR Color Gradients in Early Type Galaxies at $z \leq 1.0$}
\begin{document}

\title{Optical$-$NIR Color Gradients in Early Type Galaxies at
 $z \lesssim 1.0$\footnotemark[1]}

\footnotetext[1]{Based on observations with
 the NASA/ESA {\it Hubble Space Telescope},
 obtained at the Space Telescope Science Institute,which is 
operated by the Association of Universities for Research in Astronomy, Inc., 
under NASA contract NAS5-26555.}

\author{Sasha Hinkley\altaffilmark{2}, Myungshin Im\altaffilmark{3}}

\altaffiltext{2}{Physics Department, University of California, Santa Cruz, 
	CA 95064; shinkley@ucolick.org}
\altaffiltext{3}{Astronomy \& Astrophysics Department, University of 
	California, Santa Cruz, CA 95064; myung@ucolick.org}

\setcounter{footnote}{3}

\begin{abstract}

  Recent studies based on line indices of nearby early-type galaxies 
 suggest that stars in the galaxies' inner part may be younger 
 and more metal rich than stars in the outer part.
  If confirmed, the finding has a profound implication for the formation
 and evolution of early-type galaxies, since such age gradients naturally arise
 if merging played an important role in their formation/evolution. 
  As an indpendent test for the existence of the age gradient, 
 we investigate the optical-NIR color gradients of six field 
 early-type galaxies at $z \simeq $0.4 -- 1.0 where the age gradient is
 expected to be detectable when combined with $z=0$ calibrations. 
  By utilizing HST NICMOS H-band imaging along with
 WFPC2 optical data, we obtain a broad wavelength baseline,
 giving results that are sensitive to the expected age gradient.
  When compared with simple model predictions, five of the objects
 show negligible age gradients and are metallicity
 gradient dominated, while the remaining object shows a color 
 gradient consistent with the age+metallicity gradient picture.
  Our results suggest that stars within early-type galaxies may be coeval
 at $z \sim 1$, but sprinkling of young stars might have occured at $z < 1$.

\end{abstract}


\section{INTRODUCTION}
\label{sec:intro}

  Early-type galaxies in the local universe show a moderate 
 change in stellar populations from their centers toward their edges.
  Stars are redder toward the center of early-types (e.g., Franx \& Illingworth 1990;
 Peletier et al. 1990a, 1990b; Kormendy \& Djorgovski 1989), 
  and studies of metallicity-sensitive line indices of nearby ellipticals
 (e.g., Kobayashi \& Arimoto 1999; Gonzalez 1993; Carollo \etal 1993) show that
 a change in metallicity has a substantial effect on the color gradient.
  However, when age-sensitive Balmer line indices (see Worthey 1994) are investigated,
 some ellipticals not only show metallicity gradients, 
 but also age gradients in which the inner part is 
 younger than the outer part of the galaxy 
 (Gonzalez 1993; Tantalo, Chiosi, \& Bressan 1998; Trager et al. 2000). 

  This picture of an age+metallicity gradient makes sense 
 if mergers play a significant role in the formation/evolution
 of early-type galaxies. Only the metallicity gradient is 
 expected in models where early-type galaxies formed
 via monolithic collapse at high redshift and evolved passively thereafter
 (e.g., Carlberg 1984).
  The indication of younger stars near the center suggests additional
 star formation/accretion activities very possibly through merging.
  However, a number of studies point out difficulties of using 
 Balmer lines for age measurements (Maraston \& Thomas 2000; Bressan \etal 1996;
 Lee, Yoon, \& Lee 2001; Jorgensen 1997; Fisher, Franx, \& Illingworth 1995)      
 --- thus, an independent assessment of the age-metallicity gradient model
 is highly desired.

  One promising way to detect age gradients is to study
 color gradients of distant early-type galaxies ($z \gtrsim 0.2$). 
  As one looks back in time, younger stars 
 become more visible out of old stars where they are embedded.
  Therefore, the inner parts of distant early-type galaxies would look
 as blue as, or bluer than, the outer region if the age+metallicity gradient 
 model is correct. Consequently, the color gradient would become
 flatter or even reversed as a function of redshift.

  Color gradients of distant early-types have been studied using optical 
 HST images, and such studies have excluded models where 
 the color gradient is driven purely by age (Tamura \etal 2001;
 Saglia \etal 2000). However, they put few constraints  
 on the more plausible age+metallicity gradient model, due to the fact
 that the age effect, when buried under the metallicity gradient,
 is barely detectable in observed optical passbands (see section 5).
  Several studies find E/S0 candidates with bluer cores 
 (e.g., Abraham \etal 1999; Menanteau \etal 2001; Im \etal 2001a),
 but a detailed study of their kinematic and structural properties shows that
 these galaxies do not appear to evolve into typical E/S0s today
 (Im \etal 2001b). 

  In an attempt to test the age+metallicity gradient model,
  we have studied optical-NIR (near-infrared) color gradients of
 early-type galaxies out to $z \sim 1$,
 using HST NICMOS data for NIR and HST WFPC2 data for optical.
  In this paper, we present our 
 preliminary results which favor no significant age-gradient 
 in the majority of early-type galaxies in our sample.

\section{SAMPLE}
\label{sec:sample}

  Our sample consists of six field E/S0 galaxies at $z \lesssim 1$ of which
 five are located in the Groth strip (Rhodes \etal 2001), and one from the
 Hawaii Deep Field (Cowie \etal 1994). 
  These objects come from the NICMOS snapshot survey of distant galaxies
 (HST GO7895 program). The snapshot survey obtained HST NICMOS Camera 2
 $H$-band data (F160W, exposure time of 1200 secs per object)
 of 16 field galaxies at $0.2 \lesssim z \lesssim 1$ for which redshifts
 as well as HST 
 optical data are available in $V$(F606W) and/or $I$(F814W) bands.
 
  E/S0 galaxies are chosen using the quantitative selection
 criteria of Im \etal (2001a; 2001b) imposed on the $I$-band data. The method uses
two parameters: bulge-to-total 
 light ratio ($B/T$) which measures the significance of the bulge 
 component, and the residual parameter ($R$) which measures the smoothness
 and symmetry of the galaxy.  We adopt $B/T \geq 0.4$ and 
 $R \leq 0.08$ following Im \etal (2001a).
  Note that these parameters are derived for each galaxy in the 
 Groth strip using the 2-dimensional surface brightness fitting algoritm,
 GIM2D (Simard \etal 2001), which also returns other parameters 
 such as the half light radius ($r_{hl}$). 
  For one galaxy in the Hawaii Deep Field, the $B/T$ value is taken 
 from the MDS database (Ratnatunga, Ostrander, \& Griffiths 1999).
  Although an $R$ value is not readily available for this
 object, a visual inspection confirms that it is an E/S0. 
  The  basic data for our sample are presented in Table 1, and images are shown in
Fig.~1.
  Note that magnitudes are in the Vega system for the HST filters
 (Holtzman et al. 1995; Simard et al. 2001).


\begin{figure*}[!ht]
\psfig{figure=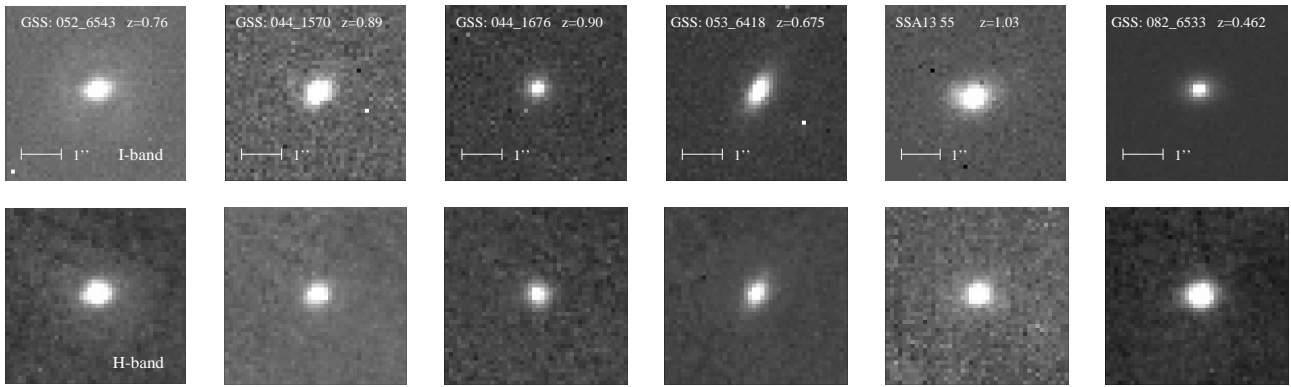,width=17.2cm}
\figcaption[hinkley_fig1.ps]{The $I$-band (top row) and the $H$-band (bottom row) images
of E/S0s in our sample. 
All images are oriented the same way and are on the same scale. }
\label{fig:images}
\end{figure*}

\section{DATA REDUCTION AND ANALYSIS}
\label{sec:dr}

  The reduction of the $H$-band data was performed using the IRAF\footnote{
 IRAF is distributed by the National Optical Astronomy Observatories,
 under contract to AURA, Inc.}
 NICPROTO package (Bushouse \etal 2000).
  The NICPROTO package\footnote{The most up-to-date NICMOS tasks
 in STSDAS now include most NICPROTO procedures.} 
 includes the standard NICMOS pipeline procedure 
 (calnica and calnicb), but it gives special attention to 
 removing image anomolies such as  a pedestal variation 
 across the array and a varying bias senstivity in the detector itself.
  In addition, all of the data were reduced using the new NICMOS synthetic
 dark frames, which helped to alleviate a time-dependent shading 
 feature caused by the readout amplifiers (Monroe \& Bergeron 1999). 
  We have carried out these additional reduction procedures in order 
 to remove the background variation as much as possible, 
 since  color gradient measurements require very careful background
 determination. 

  Four dither exposures were obtained for each field.
  From these four, a single higher signal-to-noise image was produced
 after the four were cross-correlated and shifted using subpixel translations 
with cubic spline interpolation. Simulations using tinytim PSFs confirm that  
 the simple shift-and-add method used here will not substantially change
 the characteristics of the NICMOS PSF. 
  The WFPC2 images were obtained from either the DEEP or MDS databases.
  Once the calibration and reductions were complete,
 the H-band Vega magnitudes were calculated using the 
 relation $H=21.75-2.5 \log (DN/sec)$ (Schultz \etal 2001), where
 DN (counts) is obtained from GIM2D. 

  Since NICMOS and WFPC2 have different PSFs,
 we matched
 the PSF by cross-convolving our data -- i.e., the WFPC2 images were
 convolved with the NICMOS PSF and vice versa. 
 The PSFs were created at the object location using the TinyTim package
 (Krist 1993).  Since the plate scales of the WFPC2 
 ($0 \farcs 0996$)and NICMOS Camera2 ($0\farcs 075$) are different,  
  oversampled PSFs were produced and then rebinned to have the platescale of
 the other detector.
  Note however, that no cross-convolution was performed for the $(V-I)$
 gradients since the difference in the $V$ and $I$ PSF is
 insignificant. 

 The standard IRAF photometry package ``ellipse'' was used to fit elliptical
isophotes to each object, and care was taken to ensure that the ellipticity 
and position angles were consistent between the NICMOS and WFPC2 data. Moreover, 
each object's isophotal centroid agreed extremely well between the two 
bands, and these were consistent with the GIM2D values to within 0.3 pixels.

  The background value was determined by taking a median sky 
 value of the region beyond 5 $r_{hl}$ and within a $6\farcs4$ by $6\farcs4$
 square region of the object. We checked the sky value by constructing
 the flux growth curve, and found that the growth curve nearly flattened out
 beyond $4~ r_{hl}$. The sky value is consistent with the output from
 GIM2D which uses a similar procedure ($<$10\%).
 Slight differences in sky values affect the aperture photometry
 at the outer part of the galaxy at 3 -- 5 $r_{hl}$, thus  
 we have restricted our study to regions inside
 $r \lesssim 2 r_{hl}$.
  
\section{MODELS}
\label{sec:models}

  This study uses simplified versions of the two kinds of models discussed
 in section \ref{sec:intro}. 
  The first is a pure metallicity gradient model, and the second is an 
 age+metallicity gradient model.
  The metallicity ($Z$) and the age ($t$) gradients are 
 modelled as:

\begin{eqnarray}
\delta\log(Z) & = & C + C_{Z}\log(r/r_{hl}), \\
\delta\log(t) & = & C' + C_{t}\log(r/r_{hl}),
\label{eq:metalgrad}
\end{eqnarray}

\noindent  
 where $C$ and $C'$ are constants, $r_{hl}$ is the half light radius,
 and $C_{Z}$ and $C_{t}$ are
 the slope of the metallicity gradient and the age gradient 
 respectively.

  The pure metallicity gradient model assumes no age gradient (i.e.,
 $C_{t}=0$), and local E/S0s show $C_{Z}$ values in the range of
 $-0.4 \lesssim C_{Z} \lesssim -0.1$ with $\langle C_{Z} \rangle = -0.2$ 
 (Henry \& Worthey 1999; Kobayashi \& Arimoto 1999).
  The plausible range for the central metallicity value is
 100\%--300\% $Z_{\solar}$ within $r_{hl}/8$ (Trager \etal 2001),
 and we use 250\% $Z_{\solar}$ for the inner region defined at
 $\log(\frac{r}{r_{hl}}) \simeq -1.175$
 (the light-weighted mean radius within $r_{hl}/8$). 
  The outer region is defined as where the metallicity falls to
 100\% $Z_{\solar}$. The model colors are calculated at these inner 
 and outer radii, and the model color gradient is the straight line
 which connects the two points. 
  Then, we find the best-fit $C_{Z}$ and the age (or $z_{for}$, 
the formation redshift) by minimizing a $\chi^{2}$ 
 fit to the observed color gradient.

  The age+metallicity gradient model assumes an age gradient  
 in addition to the metallicity gradient.
  Line indices studies find that $\langle C_{t} \rangle \simeq 0.1$ 
 and $\langle C_{Z} \rangle \simeq -0.25$ for local E/S0s
 (Henry \& Worthey 1999).
  Our base age+metallicity gradient model assumes these slopes,
 and the zero point of the base model is adjusted to fit
 the optical-NIR color at $r/r_{hl} \sim 0.1$ -- 0.2 by changing $z_{for}$.  
  Besides the base model, we allow our model to fit for 
 $C_{t}$, $C_{Z}$ and $z_{for}$.
  Again, 250\% $Z_{\solar}$ is placed in consideration
 for the inner metallicity.
  Note that we use only two points for the color gradient predictions.
 Consequently, the predicted color gradients are  
 straight lines.  This should be a good first order 
 approximation for the color gradients given the quality of our data,
 although the color gradients can get curved when the age effect is strong
 (e.g., Tamura et al. 2001). 



  SEDs are generated at different ages and metallicities 
 using the 1996 version of Bruzual \& Charlot model (1993). 
  The standard Salpeter IMF, 0.1 Gyr burst,
 and an open universe with $\Omega_{0}=0.2$
 and $H_{0}=70$ km sec$^{-1}$ are assumed. 
  The SEDs are redshifted and $K-$corrections are derived 
 using the formula in Im et al. (1995), which are then used for 
 calculating colors. 
  Note that the zeropoint of the optical-NIR color may vary 
 with spectral synthesis models, but the slope of the color gradient is 
 not likely to be model-sensitive (Charlot, Worthey, \& Bressan 1996). 
  Also, the predicted slope of the color gradient is insensitive to 
 the assumed cosmology. 

\section{RESULTS}
\label{sec:results}

  Color gradients for our sample are shown in Fig. 2.
  Except for one object without $V$-band data,
 two plots are shown for each object --- one for the $V-H$ gradient
 and another for the $V-I$ gradient.  
  At $z=1$, the observed $V-H$ gradient 
 corresponds roughly to the rest-frame $U-I$ gradient.
  Also plotted in Fig. 2 are
 model predictions. 


  The first five objects have $V-H$ or $I-H$ 
 gradients inconsistent with that of the base
 age+metallicity gradient model predictions,
 while their color gradients can be well matched by
 the pure metallicity gradient model with the best-fit slope of
 $C_{Z} = (-0.2\,\sim\, -0.4) \pm 0.05$.
  Fig. 2 also demonstrates that the $V-I$ gradient
 is not very sensitive to the age gradient, thus not useful for 
 testing the age-metallicity gradient model. 
   Since these objects are already at the lookback time of 
 $\sim 7$ Gyrs,  the mean stellar age in the inner part must be less than 
 a few Gyrs under the base age+metallicity 
 gradient model.
  This would produce flat or reversed optical-NIR color gradients
 (dashed lines), which does not match the data.
  Thus, it is unlikely that any significant secondary
 star formation/accretion event occured at the center of these
 galaxies before $z \sim 1$. 
  To prove this point further, we fit the color gradient for 
 $C_{t}$ using a fixed $C_{Z}=-0.25$, 
 and find no significant age gradient ($C_{t} \simeq 0.01 \pm 0.02$).
  One might consider a case where the age gradient is hidden
 under a steep metallicity gradient.  To check such possibility,
 we fit for $C_{t}$ using $C_{Z}=-0.5$ and find that
 $C_{t}\simeq 0.08 \pm 0.02$ for the five objects.
  The derived $C_{t}$ values are inconsistent with the local constraints
 where $C_{t} \simeq 0.3$ when $C_{Z} \simeq -0.5$ (Henry \& Worthey 1999).
  Thus, we still require more young stars near the center
 for this picture to be true.
  Note that the model color is sensitive to the assumed metallicity.
  If one uses a different metallicity value, the zero point of 
 the predicted color gradient shifts too.
 However, this shift in the zero-point can be compensated by
 changing $z_{for}$. We tried  100\% $Z_{\solar}$ for the metallicity
 of the central part. This changed the predicted color gradient slightly,
 but did not affect the overall conclusion about the age gradient.




  The $(V-H)$ color gradient of the last object 
 (GSS082\_6533) shows that its inner part is bluer than the outer part,
 with the best-fit gradients being $C_{Z} = -0.27 \pm 0.20$
 and $C_{t} = 0.20 \pm 0.10$.
  This is consistent with the age+metallicity gradient picture, and 
  sprinkling of young, blue stars  might have occurred 
 in the past few Gyrs, near the center of this galaxy.

  Fig. 2 also shows that the sensitivity to the age gradient seems 
 to be higher for E/S0s at high-z. For example, the difference 
 between the pure-$Z$ model and the age+metallicity gradient models
 is more drastic for the $z=0.89$ object than the $z=0.462$ object. 
  However, secondary star formation may occur after $z \sim 1$,
 in which case it is advantageous to observe galaxies at moderate $z$.



\begin{figure*}[!ht]
\psfig{figure=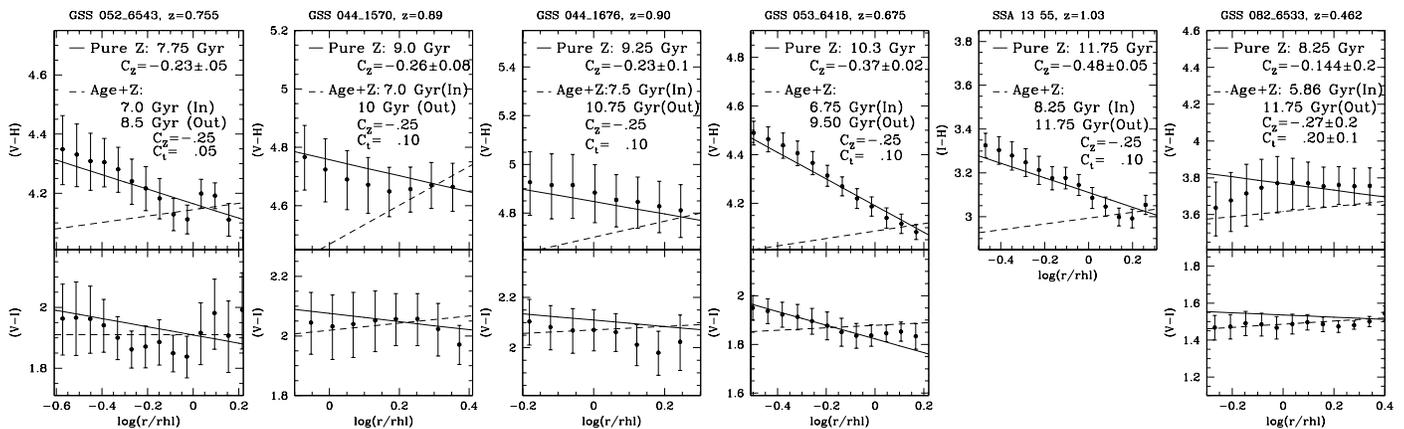,width=18.5cm,angle=-90}
\figcaption[hinkleyfig2.ps]{
 Shown here are pairs of $(V-H)$ and $(V-I)$ gradients for all 
of our data. 
Object SSA13 \#55 does not have observations in $V$, so only an $I-H$ gradient is shown. 
  The solid lines indicate the purely metallicity-driven model,
 while the dashed lines are the age + metallicity models.
  Also shown in the figure are the best-fit or assumed 
  $C_{Z}$, $C_{t}$, and the age for the models.
  For 052\_6543, a slightly shallower age gradient is used
 (0.05) since the canonical value produces a color gradient that 
 is negatively too steep.
  For a majority of this sample (5/6), the gradients follow the purely
 metallicty-driven model. Only the last object, GSS $082\_6533$,
 shows a sign of young stars in its inner region.
 In all of the plots, the points are closely spaced,  
 so two adjacent isophotes may share some pixels. Thus, 
 the error bars are not independent of each other.
  The lookback times of 6, 7, 8, 9, and 10 Gyrs correspond to
 redshifts of 0.79, 1.1, 1.5, 2.2, and 3.4 respectively. 
}  
\label{fig:cgrads1}
\end{figure*}

\section{CONCLUSIONS}
\label{sec:conc}

We have performed surface photometry on six early-type galaxies 
at $z = 0.4$ -- 1.0 and obtained their optical-NIR color gradients.
These were then compared with simple models driven purely by metallicity or
models driven by changes in age {\it and} metallicity.
The age+metallicity model predicts a flattened or reversed 
optical-NIR color gradient due to the assumed existence of younger stars
in the inner region. However, the model fails to
reproduce the observed color gradient of the majority of our galaxies
(5/6).
Therefore, we conclude that no significant secondary star
formation/accretion occured at $z \gtrsim 1$ for these galaxies. 
While evidence is lacking for a strong age-gradient
for $z \sim 1$ early-types, one object at the lowest redshift ($z=0.462$)
shows a color gradient consistent with the age+metallicity
gradient model. This shows that the age-gradient is detectable, 
and that secondary bursts/accretions may occur at moderate redshifts.  
In the future, we hope to draw firmer conclusions on this issue by  
enlarging our sample and improving our model predictions.
\acknowledgments
  This work was supported by the STScI grant GO-07895.02-96A, and   
  partial support was provided by the NSF grant AST-9529098.
  This project greatly benefitted from the help and comments provided by Luc
 Simard, Sandra Faber, David Koo, Dan Magee and Rychard Bouwens.
  We also acknowledge the use of the data from the DEEP project, which was  
 funded by the NSF grant AST-9529098, and the Medium Deep Survey,
 which was funded by STScI grant GO6951 and daughters.

\begin{deluxetable}{l c c c c c c c c c c}
\tabletypesize{\tiny}
\tablecolumns{11}
\tablecaption{Basic Data for Early-type sample}
\tablewidth{0pt}
\tablehead{Object ID & RA & DEC & $z$ & Type & $H$ & $I$ & $V$ & ${(B/T)}_I$ & $r_{hl}('')$  & $R_I$  \\
(1)            &  (2)       &  (3)       &  (4)                          &  (5) & (6)              & (7)              & (8)             & (9)              & (10)              & (11) }
\startdata
GSS$052\_6543$ & 14:17:48.2 & 52:31:17.3 & $0.756$                       & S0   &  $18.02^{+.05}_{-.05}$ & $20.31^{+.01}_{-.01}$ & $22.26^{+.04}_{-.03}$& $0.44^{+.01}_{-.02}$ & $0.637^{+.03}_{-.02}$   & $0.07$     \\
GSS$044\_1570$ & 14:18:04.8 & 52:30:59.1 & $0.89\pm0.15$\tablenotemark{*} & E/S0  & $18.96^{+.18}_{-.13}$ & $21.76^{+.04}_{-.03}$ & $23.79^{+.07}_{-.06}$& $0.71^{+.06}_{-.06}$ & $0.175^{+.019}_{-.030}$ & $0.06$        \\
GSS$044\_1676$ & 14:18:04.9 & 52:30:53.6 & $0.90\pm0.15$\tablenotemark{*} & E/S0  & $19.82^{+.10}_{-.10}$ & $22.29^{+.05}_{-.05}$ & $24.33^{+.08}_{-.10}$ & $0.56^{+.13}_{-.25}$ & $0.195^{+.061}_{-.042}$ & $0.04$       \\
GSS$053\_6418$ & 14:17:55.6 & 52:31:52.4 & $0.675$                       & Sp,S0  & $18.37^{+.07}_{-.07}$ & $20.97^{+.01}_{-.02}$ & $22.66^{+.02}_{-.03}$& $0.72^{+.03}_{-.02}$ & $0.354^{+.015}_{-.022}$ &  $0.04$      \\
SSA13 \#55     & 13:12:19.3 & 42:45:01.2 & $1.03 $                       & S0  
 & $18.53^{+.21}_{-.07}$ & $21.82^{+.22}_{-.22}$ &     $N/A$     & $0.63^{+.35}_{-.35}$ & $0.379^{+.012}_{-.012}$ &  $N/A$         \\
GSS$082\_6533$ & 14:17:28.8 & 52:27:37.8 & $0.462$                       & E/S0
 & $18.45^{+.16}_{-.26}$ & $20.57^{+.01}_{-.01}$ & $22.04^{+.01}_{-.01}$& $0.41^{+.02}_{-.02}$ & $0.207^{+.010}_{-.007}$ &  $0.05$         \\
\enddata
\label{tab:iband}
\tablecomments{ (1)  Source ID given by FFC-XXYY, where FF is the
Groth Strip subfield, C is the WFPC2 chip number, and XX and YY are the chip coordinates in units of 10 pixels. The ``SSA'' denotes that the object is
from the Hawaii Deep Field. (2)  Right Ascension (J2000). (3) Declination (J2000) (4) Spectroscopic redshifts(5) Morphological type
(6)-(8) Total magnitudes for $H$, $I$, and $V$-bands. (9)-(11)Bulge-to-total ratio, half-light radius, and
residual parameter within 2 $r_{hl}$ for the $I$-band
data as measured by GIM2D (Simard \etal 2001 in preparation).}
\tablenotetext{*}{Photometric redshifts are derived as described in Im \etal (2001a)}
\end{deluxetable}

\end{document}